\documentclass[aps,prl,twocolumn,groupedaddress,showpacs]{revtex4}
\usepackage{amsmath}
\usepackage{amsfonts}
\usepackage{color}
\usepackage{graphicx}
\usepackage{subfigure}

\def\bra#1{{\langle#1|}}
\def\ket#1{{|#1\rangle}}

\def\expectLR#1{{\left\langle#1\right\rangle}}
\def\expect#1{{\langle#1\rangle}}

\def\proj{{\hat{\cal P}}}

\def\H{{\hat H}}

\def\a{{\hat a}}
\def\adag{{\hat a}^\dagger}

\def\Op{{\hat O}}
\def\Opdag{{\hat O}^\dagger}
\def\id{{\hat I}}

\def\A{\hat{A}}
\def\Adag{\hat{A}^\dagger}

\def\Sp{\hat{\sigma}^+}
\def\Sm{\hat{\sigma}^-}

\def\projX1{{\hat{\cal P}_{X_1}}}

\def\probX1{\expect{\proj_{X_1}}}
\def\probOp{\expect{\Op}}

\begin{document}
\title{Continuous monitoring can improve indistinguishability of a single-photon source}
\author{ Shesha Raghunathan}
\email{sraghuna@usc.edu} 
\author{Todd Brun }
\email{tbrun@usc.edu}
\affiliation{Center for Quantum Information Science and Technology,
Communication Sciences Institute, Department of Electrical Engineering, 
University of Southern California
Los Angeles, CA 90089, USA.}

%
\begin{abstract}
A new engineering technique using continuous quantum measurement in conjunction with feed-forward is proposed to improve indistinguishability of a single-photon source. The technique involves continuous monitoring of the state of the emitter, processing the noisy output signal with a simple linear estimation algorithm, and feed forward to control a variable delay at the output. In the weak coupling regime, the information gained by monitoring the state of the emitter is used to reduce the time uncertainty inherent in photon emission from the source, which improves the indistinguishability of the emitted photons. 
\end{abstract}

\pacs{42.50.-p, 73.21.La, 03.65.Yz, 02.70.Ss }
\maketitle

%
Developing a scalable model to implement quantum computation on physical systems has generated consuming interest in the quantum computing community ever since Shor \cite{Shor94} and Grover \cite{Grover97} showed that quantum computing can outperform any classical device for certain algorithms of practical interest. Knill \emph{et al.} \cite{Knill01} developed a scalable model based on linear optics and single qubit measurement. This model, however, assumes the availability of good quality single photon states. This, consequently, has lead to considerable interest in the design and implementation of good quality single-photon source. The applications of a single-photon source, however, are not restricted to quantum computing alone, and would be highly useful in the areas of quantum imaging, metrology and quantum cryptography, to name a few \cite{Oxborrow05}.

Amongst various physical implementations of a single-photon source, semiconductor quantum dot (QD or dot for short) based sources are of great interest, as they scale well upon integration, and are amenable to commercial fabrication techniques \cite{Oxborrow05, QDs}. Since photon emission is isotropic in free space, spatial confinement of photons is vital to improve the collection efficiency of a single-photon source. Design and fabrication of microcavities has received much attention over the last few years, and a variety of cavities have been fabricated: micropillar, microdisk, and photonic crystal to name a few. These microcavities have different sizes and shapes, and varying quality factors \cite{Vahala03}.      

A quantum dot in a microcavity can be pumped either optically or electrically. Electrical pumping is interesting for a variety of reasons. In particular, since the electrical pumping process cannot directly insert photons into the cavity, one could pump directly into an energy level resonant with the cavity mode. Electrical pumping of a quantum dot involves the application of voltage across the source and drain; if the dot has no electron in the valence band, an electron from the source tunnels through to the dot \cite{electrical-pumping}. Once an electron tunnels through to the dot, no new electron can tunnel through due to Coulomb repulsion of like charges, a phenomenon widely referred to as the \emph{Coulomb blockade} effect \cite{QDs, electrical-pumping}. Thus, a new electron can tunnel through only when the electron currently in the dot `tunnels out.' The electron in the quantum dot tunnels out when there is a difference in voltage between the dot and the drain; this relative voltage can be controlled by the application of gate voltage on the dot. By observing the voltage across the source and drain of a quantum dot, we can also gain information about the state of the dot. This observation can be seen as a weak quantum measurement that provides some information about the system \cite{Levi07}. 

%
\begin{figure}[htp]
  \begin{center}
		 \includegraphics[width=2.75in, height=1.75in]{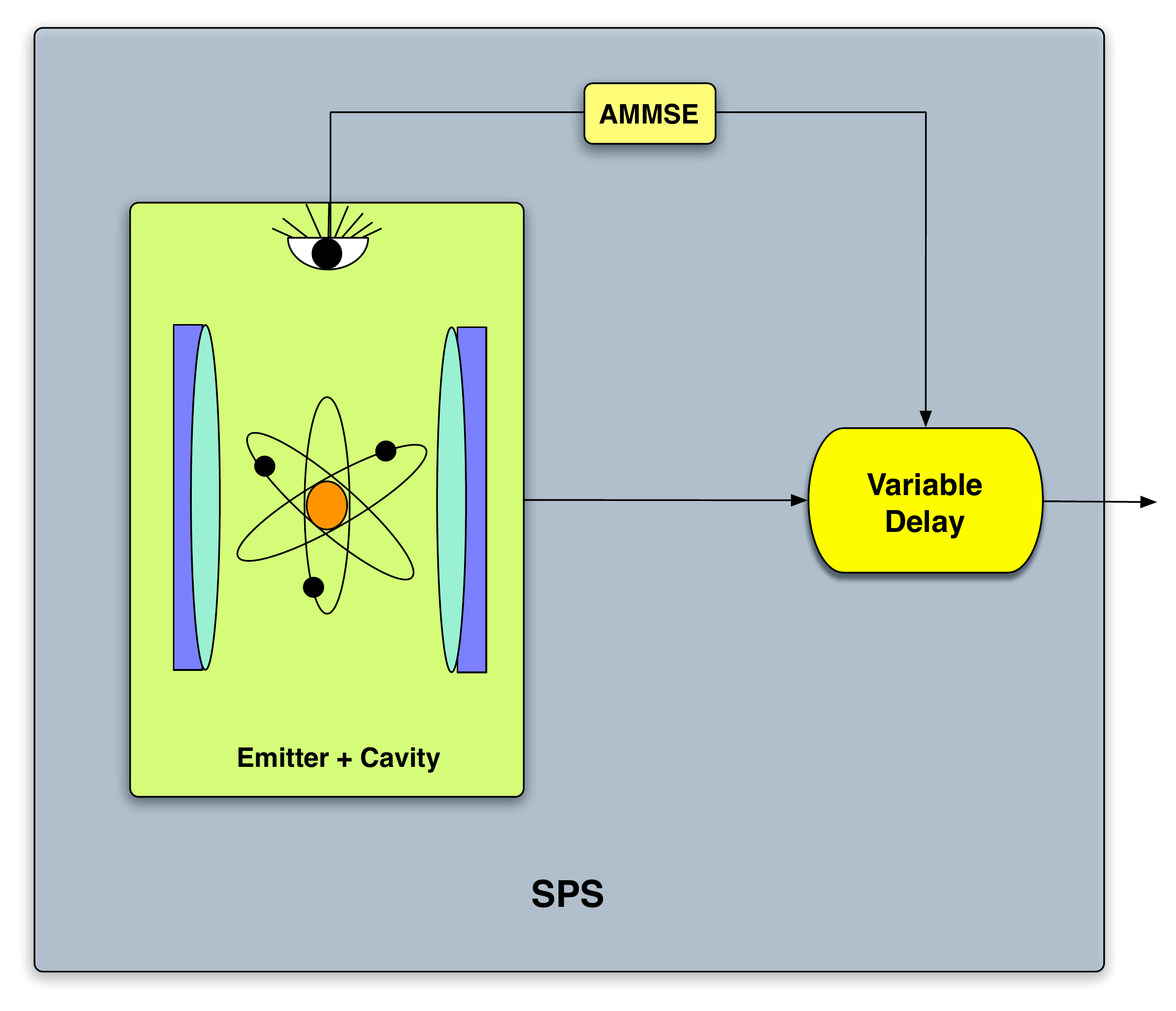}
  \end{center}
  \caption{(Color online) Continuous monitoring with feed forward involves 3 components: ($i$) emitter $+$ cavity, ($ii$) AMMSE-based transition time estimation, and, ($iii$) variable delay element. }
  \label{fig:overview}
\end{figure}
In this Letter, we combine continuous quantum measurements and feed forward to reduce the uncertainty in the time at which a photon leaks out of a source (Fig.~\ref{fig:overview}). The idea is to continuously monitor the state of the emitter, and to use this information to make a time correction at the end, so that the time uncertainty of the single-photon source is reduced. To reduce the effect of noise on our estimate of when the emitter transitioned to its ground state, we process the signal with the simple error estimation technique of Affine Minimum Mean Squared Error estimation (AMMSE). This processed signal is fed forward to control a variable delay at the output of the single-photon source. This approach works well in the weak coupling regime, as the transition time of the emitter from an excited state to its ground state controls, fairly closely, the time of the photon leaking out of the cavity. 

\begin{figure}[htp]
  \begin{center}
		 \includegraphics[width=1.75in, height=1.25in]{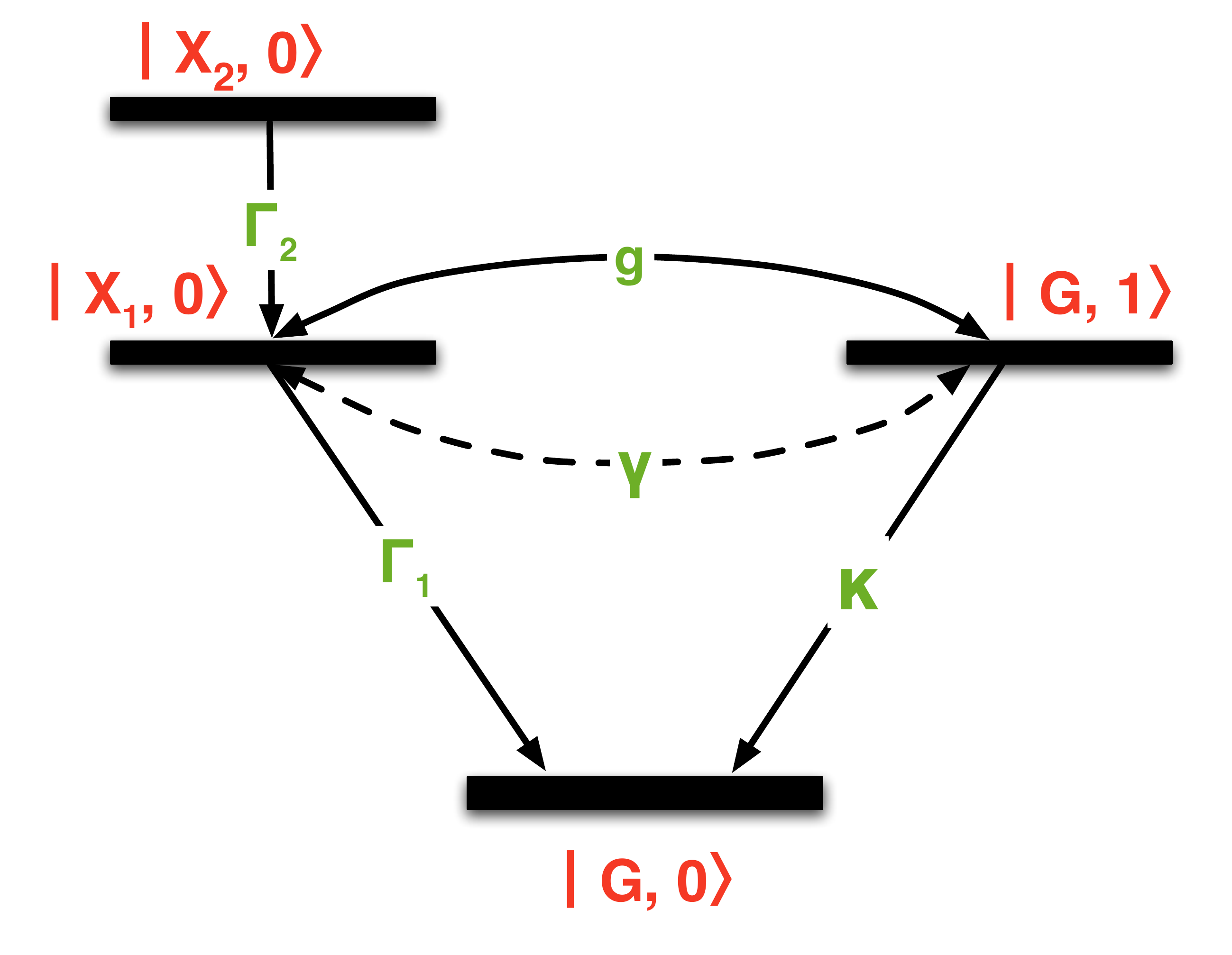}
  \end{center}
  \caption{\textsc{Emitter $+$ Cavity System:} Energy level diagram}
  \label{fig:3LA}
\end{figure}

%
Figure \ref{fig:3LA} shows the energy level diagram of the model under consideration \cite{SPS-theoretical}. The system consists of an emitter (QD) with 3 energy levels whose first excited state is resonantly coupled to the cavity mode. The energy levels are represented by $\ket{G}$, $\ket{X_1}$, $\ket{X_2}$, with $G$ standing for the ground state of the emitter while $X_1$ and $X_2$ represent the first and second excited states of the emitter, respectively. The interaction between the emitter and the cavity in the emitter $+$ cavity system is determined by the Jaynes-Cummings Hamiltonian: $ \H_I = \, i \hbar \, g \left( \adag \Sm_1 \, - \, \a \Sp_1 \right)$, where $g$ is the interaction strength, and $\Sm_1 = \ket{G}\bra{X_1}$; we operate in the interaction picture, and henceforth set the total Hamiltonian $\H$ to $\H_I$. It is also assumed that the cavity can contain at most one photon. Further, in this work, we do not model the pumping process explicitly, and we assume that the system is initially in $\ket{X_2,0}$. This assumption is reasonable if the pumping is strong.

Incoherent or decoherence processes considered in this model include spontaneous decay, cavity leakage and dephasing. $\Gamma_1$ and $\Gamma_2$ are spontaneous emission rates for $\ket{X_1} \rightarrow \ket{G}$ and $\ket{X_2} \rightarrow \ket{X_1}$ transitions respectively; $\kappa$ is the rate of (photon) leakage from the cavity, and $\gamma$ is the dephasing rate between $\ket{X_1,0}$ and $\ket{G,1}$. The time evolution of the system is described by a stochastic master equation (SME) \cite{Trajectories}:
\begin{equation}
\begin{split}
d\rho 
&= - \,  \frac{i}{\hbar} \, [ \H, \rho ] \, dt \, + \, \left( \Gamma_1\, \mathcal{H}[ \Sm_{1}]  \, + \, \Gamma_2 \,\mathcal{H}[ \Sm_{2}] \,\right) \rho \, dt \\
&  \, + \, \left( \kappa \,  \mathcal{H}[ \a] \, + \, \gamma \, \mathcal{H}[ \Op] \, \right) \rho \, dt \, + \, \sqrt{\eta \gamma} \, \mathcal{D}[\Op] \,\rho \, dW_t,
\end{split}
\label{eqn:sme}
\end{equation}
where $\mathcal{H}$ and $\mathcal{D}$ are super-operators and are defined to be $\mathcal{H}[\A] \rho = \A \rho \Adag - (\Adag \A \rho + \rho \Adag \A)/2$, and $\mathcal{D}[\Op]\rho = (\Op - \expect{\Op}) \rho + \rho (\Opdag - \expect{\Opdag})$ where $\expect{\Op}=Tr\{\Op\rho\}$. Also, $\Sm_2 = \ket{X_1}\bra{X_2}$, $\a$ is the annihilation operator acting on the cavity part of the emitter $+$ cavity system, and $\Op = \id - \ket{G}\bra{G}$ is the dephasing or ``observer'' operator acting on the emitter part of the system. Dephasing is caused by the interaction of the system (emitter) with other degrees of freedom, for instance phonon modes in the quantum dot \cite{QDs}. If we can tap into these modes, we can obtain information regarding the system of interest; $\eta$ is the efficiency with which we (the ``observer'') can read the information in these degrees of freedom. We assume that we operate in the bad cavity limit, so that the parameters satisfy the following condition:
\begin{equation}
\Gamma_1 << g, \gamma < \kappa.
\end{equation}
%

%
The output signal obtained from our continuous measurement is called the measurement record, and is given (in rescaled units) by 
\begin{equation}
J(t) = \, \probOp(t) \,  + \, \frac{\xi(t)}{\sqrt{\eta \gamma}}, 
\end{equation}
where  $\xi(t) = dW_t/dt$ is the Gaussian white noise with zero mean, $i.e.$ $\mathbb{E}[dW_t] = 0$ where $\mathbb{E}$ is the expectation or mean of a random variable, and $dW_t^2 = dt$  \cite{Trajectories}. Integrating the measurement record over time, we obtain an estimate of when the emitter transitioned to its ground state:
\begin{equation}
 \nu = \int_0^T J(t) \, dt,
 \end{equation}
where $T$ is a sufficiently long period of time. The best estimate of the transition time of the emitter would be 
\begin{equation}
 \tau = \int_0^T \probOp(t) \, dt.
 \end{equation}
We, however, have access to $J(t)$ and not $\tau$. 

To allow for this estimation error, we pass the output signal $\nu$ through an AMMSE procedure  \cite{Papoulis91}, whose output $\hat{\tau} = G \nu + m$ represents a linear estimate of the transition time of the emitter. The idea of AMMSE is to minimize the mean squared error $\mathbb{E}[|\tau - \hat{\tau}|^2]$. We derive equations for $G$ and $m$ in terms of the mean and variance of $\tau$ such that $\mathbb{E}\{|\tau - \hat{\tau}|^2\}$ is minimized:
\begin{eqnarray}
G &= &\frac{\sigma_{\tau}^2}{\sigma_{\tau}^2 \, + \, \beta^2 \, T} \ ,  \\ 
 m &=& m_{\tau} \left( \frac{\beta^2 \, T}{ \sigma_{\tau}^2 \, + \, \beta^2 \, T} \right),
\label{eqn:G&m}
\end{eqnarray}
where $m_{\tau}$ and $\sigma_{\tau}^2$ are mean and variance of signal $\tau$ repectively, $\beta  = (\eta \gamma)^{-1/2}$, and $T$ is the total integration time. In the solution above, we have, as a first approximation, assumed that there is no correlation between the output signal ($\nu$) and noise ($\nu - \tau$) affecting the system \cite{Raghunathan07}. In the bad cavity limit, the  the emitter displays a quasi-exponential behaviour \cite{Haroche06}. Since the variance of an exponential distribution is square of the mean, we have $\sigma_{\tau}^2 \approx m_{\tau}^2$.  The mean, $m_{\tau}$, is calculated by numerically integrating the deterministic (Lindblad) master equation, $i.e.$ Eq.~(\ref{eqn:sme}) with $\eta = 0.0$.

%
There are two, sometimes conflicting, requirements for an emitter to be a good single-photon source: (a) that a single photon leaks out of the source (cavity) with high probability, and, (b) that the photon that leaks out is highly indistinguishable from other photons produced by the same source. Indistinguishability is most strongly affected by the uncertainty in time of when a photon leaks out of the source. This uncertainty is unavoidable, as the the process of generating a photon involves incoherent processes; in a system represented by  Fig.~\ref{fig:3LA}, for a photon to leak out of the cavity, the system has to undergo the transitions $\ket{X_2,0} \stackrel{\Gamma_2}{\longrightarrow} \ket{X_1,0}  \stackrel{g}{\longleftrightarrow} \ket{G,1} \stackrel{\kappa}{\longrightarrow} \ket{G,0}$ where $\Gamma_2$ is an incoherent process. In the weak coupling regime, $\kappa$ dominates $g$, and hence the reabsorption rate of the energy in the cavity by the emitter (QD) is fairly small. Thus, the $\ket{X_1} \longrightarrow \ket{G}$ transition of the emitter largely determines when a photon leaks out of the cavity. This implies that knowing the state of the emitter, should reduce the time uncertainty.

\begin{figure}[htp]
  \begin{center}
    	 \includegraphics[width=3.25in, height=2.5in]{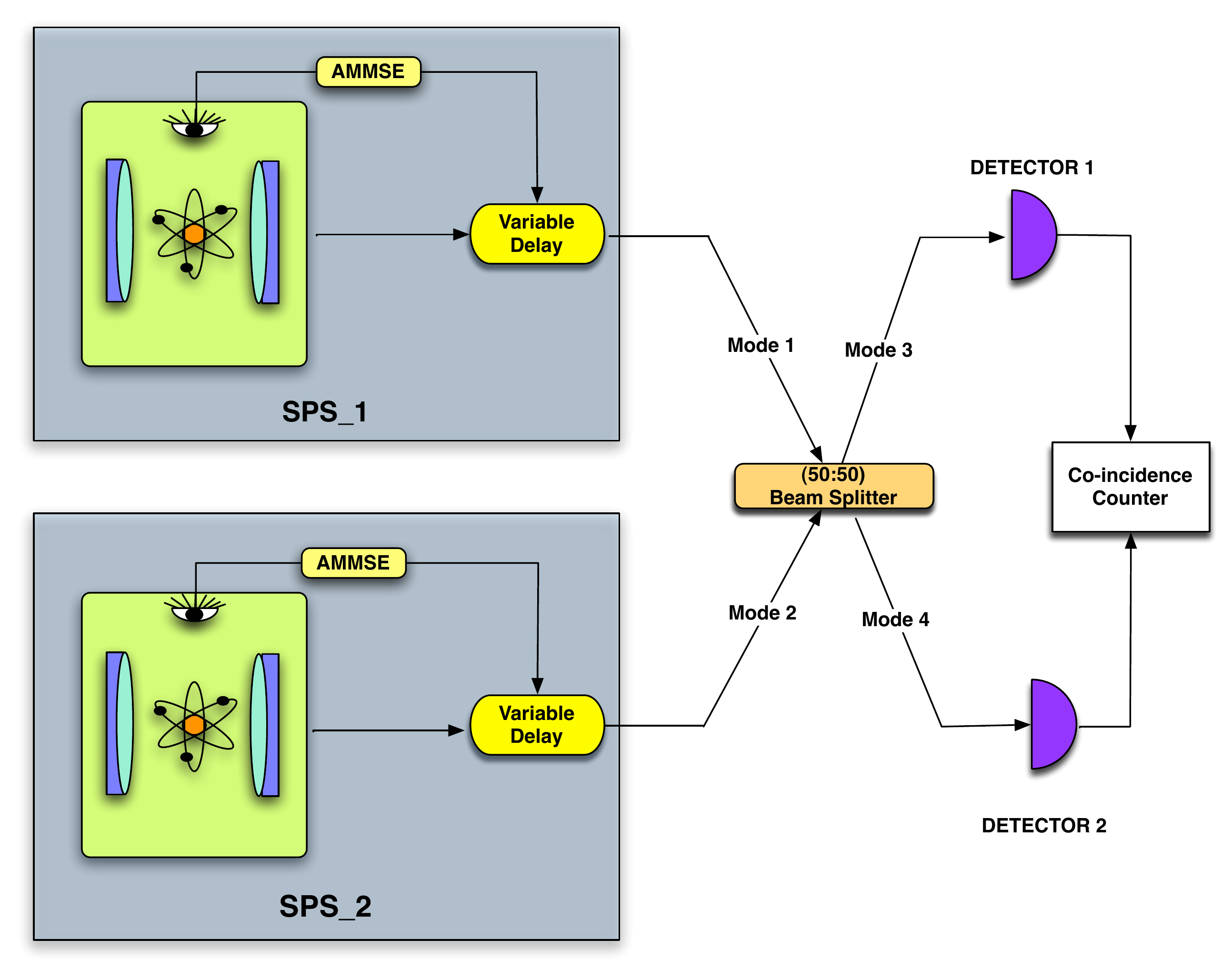}
  \end{center}
  \caption{(Color online) \textsc{Canonical experiment to calculate indistinguishability.} $SPS_1$ and $SPS_2$ are identical and independent sources with variable delays included at the output of their respective physical emitter $+$ cavity system. Variable delay is controlled by signal derived from continuous monitoring of the emitter. A ($50:50$) beam-splitter takes the photons from $SPS_1$ and $SPS_2$ as input, with output modes $3$ and $4$; coincidence counter records the output of the detectors present in output modes $3$ and $4$. If the photons in modes $1$ and $2$  are identical, they both will always go into mode $3$ or $4$.}
  \label{fig:experimental}
\end{figure}

To calculate the indistinguishability, consider a canonical Hang-Ou-Mandel-like \cite{HOM87} experimental setup as shown in Fig.~\ref{fig:experimental}. The setup has two independent, but identical, single-photon sources, $SPS_1$ and $SPS_2$;  the sources have the same parameter values $(g, \kappa, \gamma, \Gamma_2, \Gamma_1)$, and the noise acting on them is independent. The sources $SPS_1$ and $SPS_2$ have 2 components: (a) an emitter $+$ cavity system, and, (b) a variable delay. The emitter $+$ cavity system is the physical photon source, while the variable delay is introduced to correct for the uncertainty in time regarding when a photon is emitted. The delay is determined by the information gained by continuous monitoring of the system. 

Sources $SPS_1$ and $SPS_2$ emit photons into modes $1$ and $2$, respectively, which are then passed through a ($50:50$) beam-splitter whose output modes are labeled $3$ and $4$. \emph{Indistinguishability} ($\Lambda$) is defined as the lack of coincidence at the output modes of the beam-splitter, $i.e.$ $\Lambda = 1 - p_c$, where $p_c$ is the coincidence probability. The coincidence probability is the normalized second-order correlation function of the output of the beam-splitter \cite{SPS-theoretical}:
\begin{equation}
p_c = \frac{\int_0^T \, dt \, \int_0^{T-t} \, d\tau \, \, G^{(2)}_{3,4}(t,\tau) }{\int_0^T \, dt \, \int_0^{T-t} \, d\tau \, \expectLR{\adag_3(t) \, \a_3(t)} \,  \expectLR{\adag_4(t+\tau) \, \a_4(t+\tau)}}
\end{equation}
where $G^{(2)}_{3,4}(t,\tau) = \expectLR{\adag_3(t) \, \adag_4(t+\tau) \, \a_4(t+\tau) \, \a_3(t)}$ is the second order correlation function between output modes $3$ and $4$. The output modes of the beam-splitter can be expressed in terms of its input modes using simple linear equations $\a_3(t) \, = ( \a_1(t) \, - \, \a_2(t) )/\sqrt{2}$ and $\a_4(t) \, = \, ( \a_1(t) \, + \, \a_2(t) )/\sqrt{2}$. Assuming that the photons from sources $SPS_1$ and $SPS_2$ do not scatter into modes other than $1$ and $2$, modes $1$ and $2$ are proportional to the cavity mode in their respective sources ($SPS_1$ and $SPS_2$).
%

%
\begin{figure}[htp]
  \begin{center}
    	 \includegraphics[width=3.25in,height=3.25in]{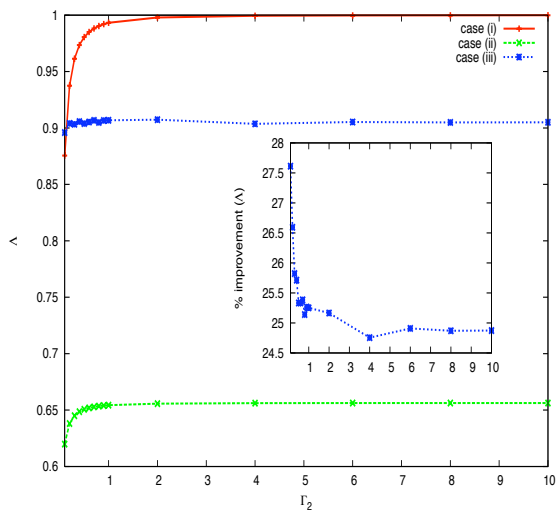}
  \end{center}
  \caption{(Color online) \textsc{Results.}  We plot indistinguishability ($\Lambda$) as a function of $\Gamma_2$ with system parameters: $g=0.1, \kappa=1, \gamma=0.1, \Gamma_1=0.001$. The 3 cases are: ($i$) no dephasing, ($ii$) dephasing but no feed forward, and, ($iii$) dephasing and with feed forward. (Inset) We plot $\%$ improvement in $\Lambda$ of case~($iii$) with respect to case~($ii$).}
  \label{fig:results}
\end{figure}
To analyze the performance of the continuous monitoring feed forward technique, we compare 3 cases: ($i$) no dephasing, ($ii$) dephasing but no feed forward correction, and, ($iii$) dephasing with feed forward correction. In case~($i$), the evolution of the system is given by a deterministic (Lindblad) master equation (Eq.~(\ref{eqn:sme}) with $\gamma=0$). In case~($ii$), there is dephasing in the system ($\gamma \ne 0$), but we (the ``observer'') either have no access to the information in the external modes coupled to the system, or, we choose to ignore the information; the evolution of this system is deterministic as well (Eq.~(\ref{eqn:sme}) with $\eta=0$). Case~($iii$) is the continuous monitoring feed forward technique, which reduces uncertainty in time by introducing a variable delay at the output of the source.

We simulated this system numerically using fourth-order Runge-Kutta integrator  ($rk4$) along with a pseudo-random number generator ($RNG$) \cite{Numerics}; the integrator was used to evolve the stochastic master equation, while $RNG$ was used to generate Gaussian white noise ($dW_t$ in Eq.~(\ref{eqn:sme})). Monte-Carlo simulation of the SME (Eq.~(\ref{eqn:sme})) was carried out by averaging $7000$ trajectories (case ($iii$) in Fig.~\ref{fig:results}).

In Fig.~\ref{fig:results}, we plot indistinguishability ($\Lambda$) as a function of $\Gamma_2$ for the 3 cases considered above. The plot is for $g=0.1, \kappa=1, \gamma=0.1, \Gamma_1=0.001$ and for case~($iii$) we have set $\eta=1$. It can be seen that $\Lambda$ for case~($i$) performs the best  while case~($ii$) performs the worst. However, we find that continuous monitoring based feed forward technique performs significantly better than case~($ii$). From Fig.~\ref{fig:results} we see that the performance improvement as compared with dephasing with no feed forward is about $25\%$ or more. The enhancement is larger for smaller $\Gamma_2$; this is due to the fact that for smaller values of $\Gamma_2$, the time uncertainty of when a photon leaks out is higher, and the impact of feed forward based time-correction (due to variable delay) is larger.

%
In the numerical study presented above, it was assumed that we get information with very high efficiency, $i.e.$ $\eta = 1$, and that the process of continuous monitoring does not in itself increase dephasing. Both these assumptions need not be true in reality. One of the motivations for developing this technique is that the same circuit used in pumping can provide information regarding the state of the quantum dot. In this case, the continuous monitoring should not have an adverse impact on the system. However, if other techniques have to be used for pumping and continuous monitoring, then it is possible that monitoring may increase dephasing in the system. Engineering a quantum dot as a single-photon source requires a more detailed study of such adverse effects, and a wider parameter space needs to be explored \cite{Raghunathan07}. 

That said, what this work shows is that there is potentially a lot to be gained by continuous monitoring (in conjuction with feed forward) to minimize the time uncertainty of photons from a single-photon source, and that this approach is simple enough to be implemented with current technology.

%
In conclusion, we have shown that continuous monitoring can be used to improve the indistinguishability of a single-photon source. The technique follows a process of (a) continuously monitoring the state of the emitter, (b) processing the noisy output, and (c) feeding forward the information gained to a variable delay at the output of the single-photon source. This simple approach led to a significant improvement in indistinguishability (about $25\%$ or more) in numerical simulations; the most encouraging fact is that this approach requires only a simple linear algorithm along with variable delay elements, which should be achievable with current technology. We believe that continuous monitoring with feed forward has the potential to be a practical engineering tool to improve performance, and similar approaches will find more applications in future.

%
{\bf Acknowledgements.} We thank A.F.J. Levi for useful discussions regarding implementation of continuous monitoring in QDs. This work was supported in part by NSF Grant No.~ECS-0507270 and NSF CAREER Grant No.~CCF-0448658.

%

\end{document}